\documentclass[prl,aps,twocolumn,superscriptaddress,showpacs,showkeys,amsmath,amssymb,floatfix]{revtex4}
\usepackage[colorlinks=true,citecolor=blue,filecolor=blue,linkcolor=blue,urlcolor=blue,pdftex]{hyperref}

\usepackage{xspace} 
\usepackage[usenames,dvipsnames]{color}
\usepackage{booktabs,graphicx,mathrsfs,verbatim,amsmath,units,soul}
\usepackage{xspace} 
\usepackage{upgreek}
\usepackage{gensymb}
\usepackage{amsmath}
\usepackage{tabularx}

\newcommand{\ton}{t}
\newcommand{\nxenontotalmass}{$3.2$\;\ton\xspace}
\newcommand{\nxenontargetmass}{$2.00$\;\ton\xspace}
\newcommand{\ntpcdriftlength}{$97~\mathrm{cm}$\xspace}
\newcommand{\ntpcdiameter}{$96~\mathrm{cm}$\xspace}

\newcommand{\nerrejectionpct}{99.7\%\xspace}

\newcommand{\nxenonnttargetmass}{5.9\;\ton\xspace}

\newcommand{\nliquidleveltolerance}{2\%~RMS\xspace}
\newcommand{\nliquidlevelabovegate}{$2.5~\mathrm{mm}$\xspace}
\newcommand{\nxenoninternaltemp}{$-96.0$\,$^\circ$C\xspace}
\newcommand{\nxenoninternaltemptolerance}{0.02\%~RMS\xspace}
\newcommand{\nxenonpressure}{1.94~bar\xspace}

\newcommand{\nlargeststwoxycorrection}{32\%\xspace}
\newcommand{\driftfieldsrzero}{$120$~V/cm\xspace} 
\newcommand{\driftfieldsrone}{$81$~V/cm\xspace} 
\newcommand{\driftfielderror}{2.2~V/cm RMS\xspace} 

\newcommand{\SRzero}{SR0\xspace}
\newcommand{\SRone}{SR1\xspace}
\newcommand{\nlivetimesrzero}{32.1~days\xspace}
\newcommand{\nlivetimesrone}{246.7~days\xspace}
\newcommand{\nlivetimetotal}{278.8~days\xspace}
\newcommand{\nstartofsciencerunzero}{November~22,~2016\xspace}
\newcommand{\nendofsciencerunone}{February~8,~2018\xspace}

\newcommand{\ndaqdeadtimesrzero}{7.8\%\xspace}

\newcommand{\ndaqdeadtimesrone}{1.2\%\xspace}
\newcommand{\nearthquakedate}{January~18,~2017\xspace}
\newcommand{\nsronestart}{February~2,~2017\xspace}

\newcommand{\nblindedpmts}{36\xspace}

\newcommand{\nlossmuon}{1.2\%\xspace}
\newcommand{\nlosstails}{4.4\%\xspace}
\newcommand{\ndaysrn}{17.1 days\xspace}
\newcommand{\ndaysambe}{30.0 days\xspace}
\newcommand{\ndaysng}{1.9 days\xspace}

\newcommand{\ncoremassname}{0.65~\ton\xspace}
\newcommand{\nonetonnemassname}{0.9~\ton\xspace}
\newcommand{\nnominalfiducialmassname}{1.3~\ton\xspace}
\newcommand{\nnominalfiducialmass}{$(1.30~\pm~0.01)$\;\ton\xspace}
\newcommand{\nexposure}{1.0\;\ton$\times$yr\xspace}

\newcommand{\nkrlevel}{$^\mathrm{nat}\mathrm{Kr}/\mathrm{Xe} = (0.66\pm0.11)~\mathrm{ppt}$}
\newcommand{\krrate}{$(7.7\pm 1.3)$~events/$(\mathrm{\ton}\times\mathrm{yr}\times\kevee)$}

\newcommand{\nlowestbg}{$(82\substack{+5 \\ -3}\textrm{~(sys)}\pm3\textrm{~(stat)})$~\dru\xspace}

\newcommand{\rntwo}{${}^{222}$Rn\xspace}
\newcommand{\rnzero}{${}^{220}$Rn\xspace}

\newcommand{\krm}{${}^{83\mathrm{m}}$Kr\xspace}
\newcommand{\ambe}{${}^{241}$AmBe\xspace} 

\newcommand{\cSone}{$\mathrm{cS1}$\xspace}
\newcommand{\cStwo}{$\mathrm{cS2}$\xspace}
\newcommand{\cStwob}{$\mathrm{cS2_b}$\xspace}

\newcommand{\z}{Z\xspace}
\newcommand{\radius}{R\xspace}
\newcommand{\radiussquared}{R$^{2}$\xspace}

\newcommand{\nmoststringentlimit}{$4.1\times10^{-47}$~cm$^2$\xspace}
\newcommand{\nmoststingentmass}{30~\gevcsq}

\newcommand{\nbestfitsigmasitwoh}{$4.7\times10^{-47}$~cm$^2$\xspace}

\newcommand\figref[1]{Fig.~\ref{#1}}
\newcommand\Figref[1]{Figure~\ref{#1}}
\newcommand\tabref[1]{Table~\ref{#1}}
\newcommand\figsref[2]{Figs.~\ref{#1}~and~\ref{#2}}
\newcommand\Figsref[2]{Figures~\ref{#1}~and~\ref{#2}}

\newcommand{\kevee}{\mathrm{keV_{ee}}}
\newcommand{\kevnr}{\mathrm{keV_{nr}}}

\newcommand{\AC}{AC\xspace}
\newcommand{\FM}{fiducial mass\xspace}

\newcommand{\dru}{events/$(\mathrm{\ton}\times\mathrm{yr}\times\kevee)$}
\newcommand{\gevcsq}{GeV/c${}^2$\xspace}
\newcommand{\tevcsq}{TeV/c${}^2$\xspace}

\newcommand{\searchregionkev}{[1.4,~10.6]~$\kevee$\xspace}
\newcommand{\searchregionkevnr}{[4.9,~40.9]~$\kevnr$\xspace}



\newcommand{\bologna}{\affiliation{Department of Physics and Astronomy, University of Bologna and INFN-Bologna, 40126 Bologna, Italy}}

\newcommand{\chicago}{\affiliation{Department of Physics \& Kavli Institute for Cosmological Physics, University of Chicago, Chicago, IL 60637, USA}}

\newcommand{\coimbra}{\affiliation{LIBPhys, Department of Physics, University of Coimbra, 3004-516 Coimbra, Portugal}}

\newcommand{\columbia}{\affiliation{Physics Department, Columbia University, New York, NY 10027, USA}}

\newcommand{\lngs}{\affiliation{INFN-Laboratori Nazionali del Gran Sasso and Gran Sasso Science Institute, 67100 L'Aquila, Italy}}

\newcommand{\mainz}{\affiliation{Institut f\"ur Physik \& Exzellenzcluster PRISMA, Johannes Gutenberg-Universit\"at Mainz, 55099 Mainz, Germany}}

\newcommand{\heidelberg}{\affiliation{Max-Planck-Institut f\"ur Kernphysik, 69117 Heidelberg, Germany}}

\newcommand{\munster}{\affiliation{Institut f\"ur Kernphysik, Westf\"alische Wilhelms-Universit\"at M\"unster, 48149 M\"unster, Germany}}

\newcommand{\nikhef}{\affiliation{Nikhef and the University of Amsterdam, Science Park, 1098XG Amsterdam, Netherlands}}

\newcommand{\nyuad}{\affiliation{New York University Abu Dhabi, PO Box 129188, Abu Dhabi, United Arab Emirates}}

\newcommand{\purdue}{\affiliation{Department of Physics and Astronomy, Purdue University, West Lafayette, IN 47907, USA}}

\newcommand{\rpi}{\affiliation{Department of Physics, Applied Physics and Astronomy, Rensselaer Polytechnic Institute, Troy, NY 12180, USA}}

\newcommand{\rice}{\affiliation{Department of Physics and Astronomy, Rice University, Houston, TX 77005, USA}}

\newcommand{\stockholm}{\affiliation{Oskar Klein Centre, Department of Physics, Stockholm University, AlbaNova, Stockholm SE-10691, Sweden}}

\newcommand{\subatech}{\affiliation{SUBATECH, IMT Atlantique, CNRS/IN2P3, Universit\'e de Nantes, Nantes 44307, France}}

\newcommand{\torino}{\affiliation{INFN-Torino and Osservatorio Astrofisico di Torino, 10125 Torino, Italy}}

\newcommand{\ucla}{\affiliation{Physics \& Astronomy Department, University of California, Los Angeles, CA 90095, USA}}

\newcommand{\ucsd}{\affiliation{Department of Physics, University of California, San Diego, CA 92093, USA}}

\newcommand{\wis}{\affiliation{Department of Particle Physics and Astrophysics, Weizmann Institute of Science, Rehovot 7610001, Israel}}

\newcommand{\zurich}{\affiliation{Physik-Institut, University of Zurich, 8057  Zurich, Switzerland}}

\newcommand{\paris}{\affiliation{LPNHE, Universit\'{e} Pierre et Marie Curie, Universit\'{e} Paris Diderot, CNRS/IN2P3, Paris 75252, France}}

\newcommand{\freiburg}{\affiliation{Physikalisches Institut, Universit\"at Freiburg, 79104 Freiburg, Germany}}

\begin{document}

\title{Dark Matter Search Results from a One Tonne$\times$Year Exposure of XENON1T}

\author{E.~Aprile}\columbia
\author{J.~Aalbers}\nikhef
\author{F.~Agostini}\bologna
\author{M.~Alfonsi}\mainz
\author{L.~Althueser}\munster
\author{F.~D.~Amaro}\coimbra
\author{M.~Anthony}\columbia
\author{F.~Arneodo}\nyuad
\author{L.~Baudis}\zurich
\author{B.~Bauermeister}\stockholm
% ...
\collaboration{XENON Collaboration}
\email[]{xenon@lngs.infn.it}
\noaffiliation

\date{\today} 

\begin{abstract}
We report on a search for Weakly Interacting 
Massive Particles (WIMPs) using \nlivetimetotal  
 of data collected with the XENON1T experiment at LNGS. 
XENON1T utilizes a liquid xenon time projection chamber with a fiducial mass of 
\nnominalfiducialmass, resulting in a \nexposure exposure.
The energy region of interest, \searchregionkev (\searchregionkevnr), exhibits an ultra-low electron recoil background rate of \nlowestbg. 
No significant excess over background is found and a profile likelihood analysis parameterized 
in spatial and energy dimensions excludes new parameter space for the WIMP-nucleon spin-independent elastic scatter cross-section for WIMP masses above 6~\gevcsq, with a minimum of \nmoststringentlimit
at \nmoststingentmass and 90\% confidence level. 

\end{abstract}



\maketitle


An abundance of astrophysical observations suggests the existence of a non-luminous,
massive component of the universe called dark matter (DM)~\cite{wimp_hooper, wimp_review}. The Weakly Interacting Massive Particle (WIMP) is one of the most 
promising DM candidates, motivating 
numerous terrestrial and astronomical searches~\cite{teresa_review,indirect}.
The most successful class of direct detection experiments searching for WIMPs with masses between a 
few \gevcsq to \tevcsq 
have utilized liquid 
xenon (LXe) time projection chambers (TPCs) and set stringent limits on the 
coupling of WIMPs to matter, excluding the WIMP-nucleon spin-independent elastic cross-section, $\sigma_{SI}$, for a 
30~\gevcsq WIMP to below $10^{-46}~\mathrm{cm}^{2}$~\cite{xe_sr0, lux_dm, panda_dm}. 

The XENON1T experiment~\cite{xe_instrument}, located at an average depth of 3600~m water-equivalent at
the INFN Laboratori Nazionali del Gran Sasso (LNGS), is the largest such detector
to date containing \nxenontotalmass of ultra-pure LXe with 2~\ton~employed as the target material
 in the active volume. This PTFE-lined, \ntpcdiameter diameter cylinder is instrumented 
above and below by arrays of 127 and 121 Hamamatsu R11410-21 3"
photomultiplier tubes (PMTs)~\cite{xe_pmts, pmt2}. 
A particle incident on the LXe target deposits energy 
that produces a prompt scintillation signal (S1) and ionization electrons. 
The active volume is defined by a cathode and 
a grounded gate electrode separated by \ntpcdriftlength to provide a drift field for the electrons. 
These electrons are extracted into gaseous xenon (GXe) where they produce 
proportional scintillation light (S2) via electroluminescence through 
a $\gtrsim10$~kV/cm multiplication field. 
The S2/S1 size ratio allows for discrimination between
nuclear recoils (NRs; from WIMPs or neutrons) and electronic recoils (ERs; from $\beta$ or $\gamma$).
The time delay between S1 and S2 and the localization of the S2 pattern in the top PMT array indicate 
the vertical and horizontal position of the interaction, respectively. 
The detector is surrounded by an active water Cherenkov muon veto system~\cite{xe_muonveto}. 


This DM search combines data from two science runs which spanned from \nstartofsciencerunzero to \nearthquakedate (\SRzero~\cite{xe_sr0}, re-analyzed in this work) and \nsronestart to \nendofsciencerunone (\SRone), with the brief interruption due to an earthquake. 
The livetime is reduced by \ndaqdeadtimesrzero (\ndaqdeadtimesrone) for \SRzero (\SRone) 
when the data acquisition system was insensitive to new events, 
\nlossmuon when the muon veto was either disabled or triggered in coincidence with the TPC, 
and \nlosstails after high-energy events in the TPC that induced photo-ionization and delayed electron extraction activity~\cite{sorensen}, resulting in \nlivetimesrzero and \nlivetimesrone for \SRzero and \SRone, respectively.
The two science runs differ in the cathode voltage
of $-12$~kV (\SRzero) and $-8$~kV (\SRone), 
corresponding to drift fields of \driftfieldsrzero and \driftfieldsrone,
with position-dependent variations of \driftfielderror
based on a field map derived with the \textsc{KEMField} simulation package~\cite{kemfield} and cross-checked by a data-driven method.
The LXe level is maintained at \nliquidlevelabovegate above the gate electrode, within sensor reading fluctuations of \nliquidleveltolerance. 
The LXe temperature and GXe pressure were constant at \nxenoninternaltemp
and \nxenonpressure, both with $<$~\nxenoninternaltemptolerance. For this analysis, 
\nblindedpmts~PMTs are ignored 
due to vacuum leaks
or low single photo-electron (SPE) acceptance~\cite{xe_instrument}.

Several internal and external radioactive sources were deployed to calibrate the detector. 
\krm calibration data~\cite{kr} were collected every $\sim$2.5 weeks  
to monitor various detector parameters.
Low-energy ERs are calibrated with \ndaysrn of data taken with 
an internal \rnzero source~\cite{xe_rn220},
split into seven periods spread throughout the science runs. 
NR calibration is performed with \ndaysambe of 
exposure to an external \ambe source,
roughly split between science runs, and \ndaysng of exposure 
to a D-D neutron generator~\cite{xe1t_ng} in \SRone.


Each PMT channel is continuously digitized
at a rate of 100~MHz. The PMT gains range from $(1-5)\times10^{6}$ and signals above a minimum 
threshold of 2.06~mV are recorded as ``hits''
resulting in a mean SPE acceptance of 93\% with a standard deviation of 3\% across all
active channels. A software trigger searches in time for clusters of hits
compatible with S1 or S2 signals and saves the duration corresponding to the
maximum drift time ($\sim$700$~\mu$s) around them. 
This data is simultaneously backed up to tape and transferred to external grid sites where it is processed~\cite{pearc_computing} with the PAX reconstruction software package~\cite{xe_instrument, pax}. 

PMT signals are corrected for time-dependent gains. 
The gains are monitored weekly with a pulsed LED configured to produce signals of a few photoelectrons (PE)~\cite{pmt_calibration} and are stable
within 1-2\% throughout each science run, except in a small number of PMTs whose voltages were intentionally lowered due to diffusive leaks or which experienced dynode deterioration. S1 signals are corrected (\cSone)
for position-dependent light collection efficiency (LCE) due to geometric effects, measured in
\krm calibration data and showing a maximum deviation of 80\%. 
While drifting through LXe, electrons can be captured by electronegative impurities and thus the S2 size must be corrected for electron lifetime, which is measured with high time-granularity using $\alpha$-decays
from \rntwo daughters during DM search data-taking as well as with \krm calibration data.
The electron lifetime increased from $380~\mu$s at the beginning of \SRzero to a plateau of 
$\sim$650~$\mu$s at the end of \SRone due to decreased outgassing over time and continuous GXe purification through hot getters.
S2 signals are additionally corrected (\cStwo) for position-dependent LCE
and inhomogeneous electroluminescence amplification,
a maximum effect of
\nlargeststwoxycorrection from the edge of the TPC to center.
The bottom PMT array (\cStwob) is used for S2 energy reconstruction due to a more spatially homogeneous LCE.

\begin{figure}[htbp]
\centering
\includegraphics[width=\columnwidth]{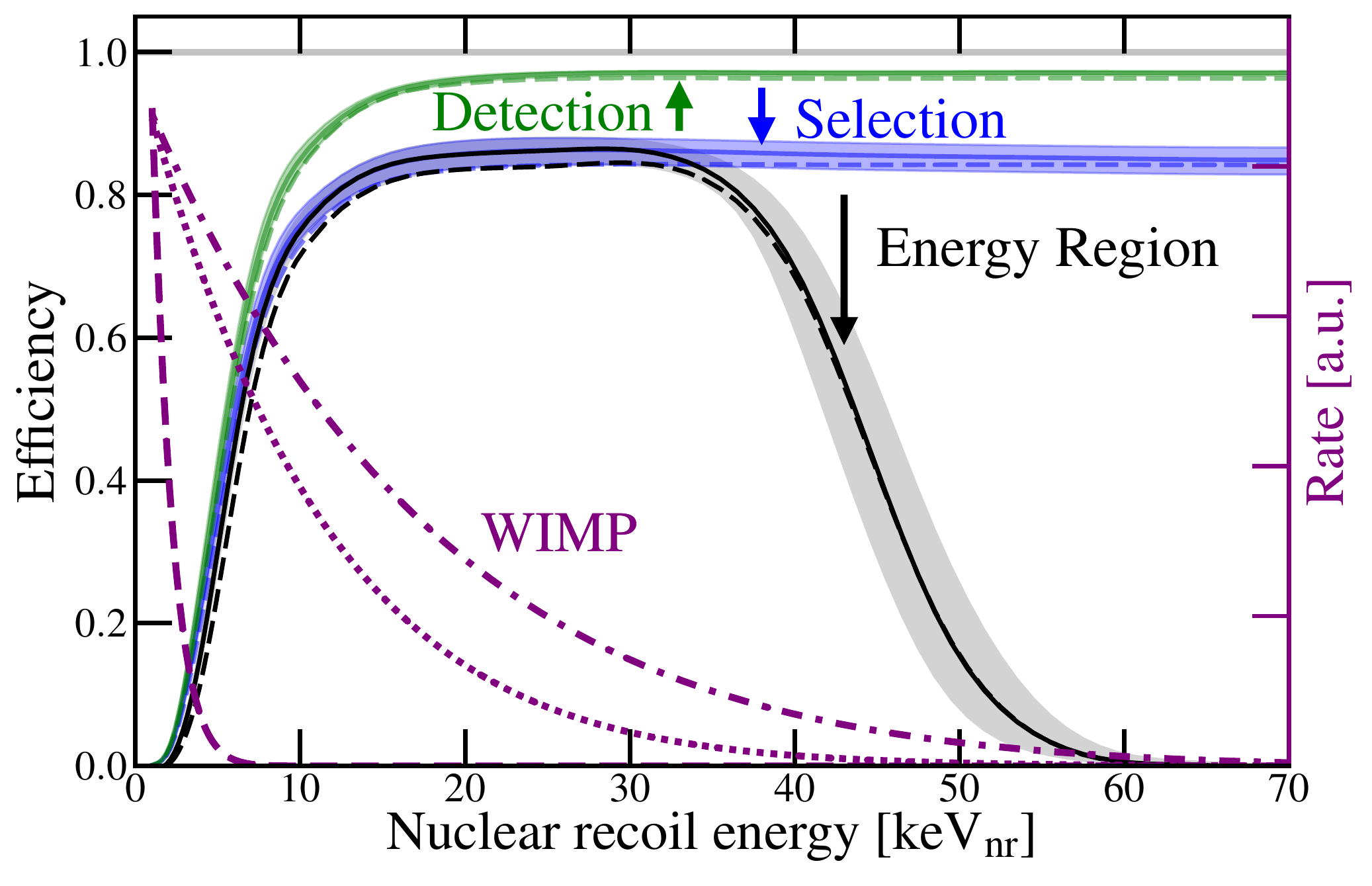}
\caption{\label{fig:acc}Best-fit total efficiencies (black), including the energy ROI selection, for \SRzero (dashed) and \SRone (solid) as a function of true NR energy ($\kevnr$). The efficiency of S1 detection (green) and that of S1 detection and selection (blue) are shown. The shaded bands show the 68\% credible regions for \SRone. The expected spectral shapes (purple) of 10~\gevcsq (dashed), 50~\gevcsq (dotted), and 200~\gevcsq (dashed dotted) WIMPs are overlaid for reference. 
}
\end{figure}

Position reconstruction in the horizontal plane employs an artificial neural network  trained with 
simulated S2 top-array PMT
hit-patterns. This Monte Carlo (MC) simulation includes the
full detector geometry, optical photon propagation, PMT quantum efficiencies, multiple-PE emission by one photon~\cite{double_pe}, and gains. 
Optical parameters are tuned to match the S1 LCE and the fraction of the S2 signal in the top-array in \krm data.
Drift field distortion causes an inward shift of the reconstructed position
and is corrected using \krm data
to obtain the horizontal (X and Y, giving radius, \radius) 
and vertical (\z) interaction positions. The bottom of the TPC (\z$=-96.9$~cm) shows the
largest radial bias of 7.7~cm (12.2~cm) at the beginning (end) of DM search data taking,
with time-dependence mostly due to gradual charge accumulation on the PTFE surfaces, similar to the observation by~\cite{lux_field}. 
The resulting position distributions 
of both spatially homogeneous \rntwo-chain $\alpha$-decays 
and $^{\mathrm{131m}}$Xe decays, 
as well as localized NRs from external \ambe and neutron generator calibration data, 
agree well with MC and validate this correction procedure.

\begin{figure*}[t]
\centering
\includegraphics[height=200pt]{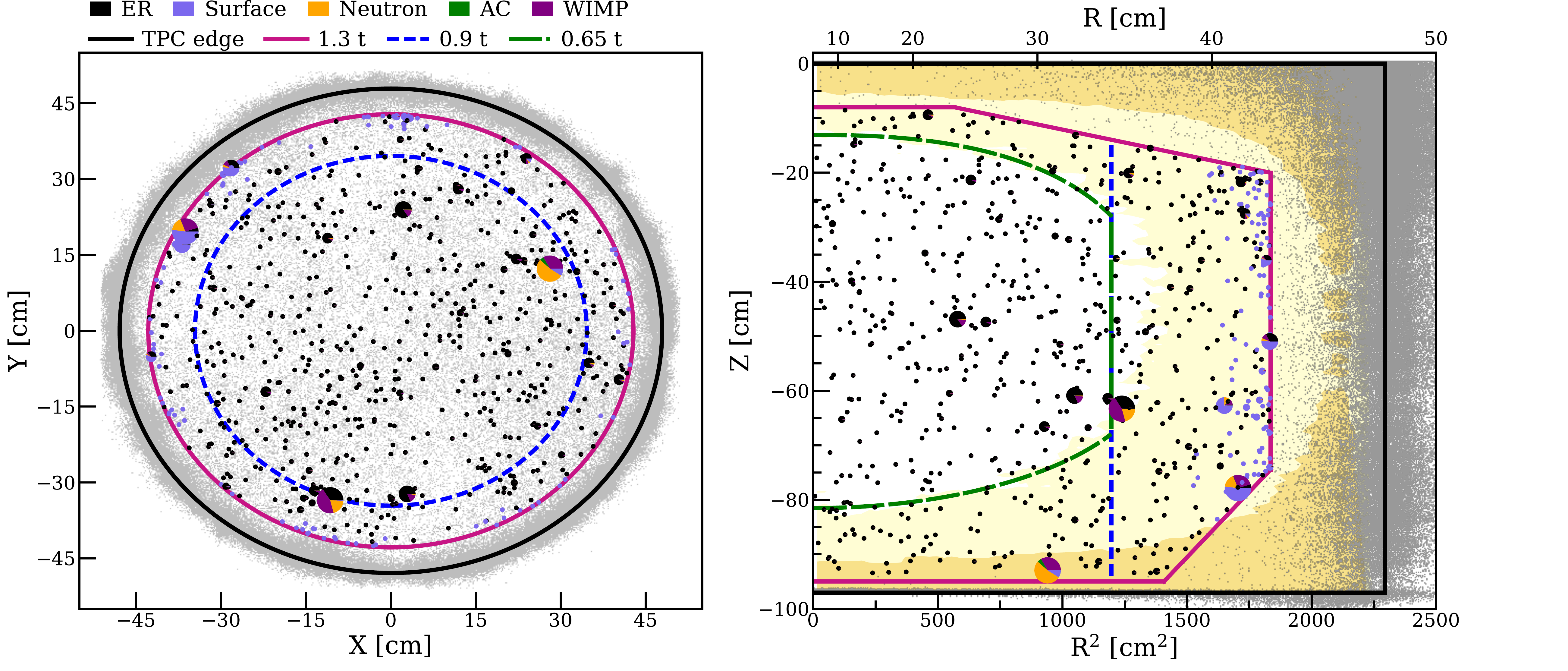}
\vspace{-0.1in}
\caption{\label{fig:position_plots}Spatial distributions of DM search data. Events that pass all selection criteria and are within the \FM are drawn as pie charts representing the relative probabilities of the background and signal components for each event under the best-fit model (assuming a 200~\gevcsq WIMP and resulting best-fit $\sigma_{SI}$~=~\nbestfitsigmasitwoh) with color code given in the legend. Small charts (mainly single-colored) correspond to unambiguously background-like events, while events with larger WIMP probability are drawn progressively larger. Gray points are events reconstructed outside the \FM. The TPC boundary (black line), \nnominalfiducialmassname \FM (magenta), maximum radius of the reference \nonetonnemassname mass (blue dashed), and \ncoremassname core mass (green dashed) are shown. Yellow shaded regions display the $1\sigma$ (dark), and $2\sigma$ (light) probability density percentiles of the radiogenic neutron background component for \SRone. 
}
\end{figure*}

\begin{figure*}[htbp]
\centering
\includegraphics[width=0.8\textwidth]{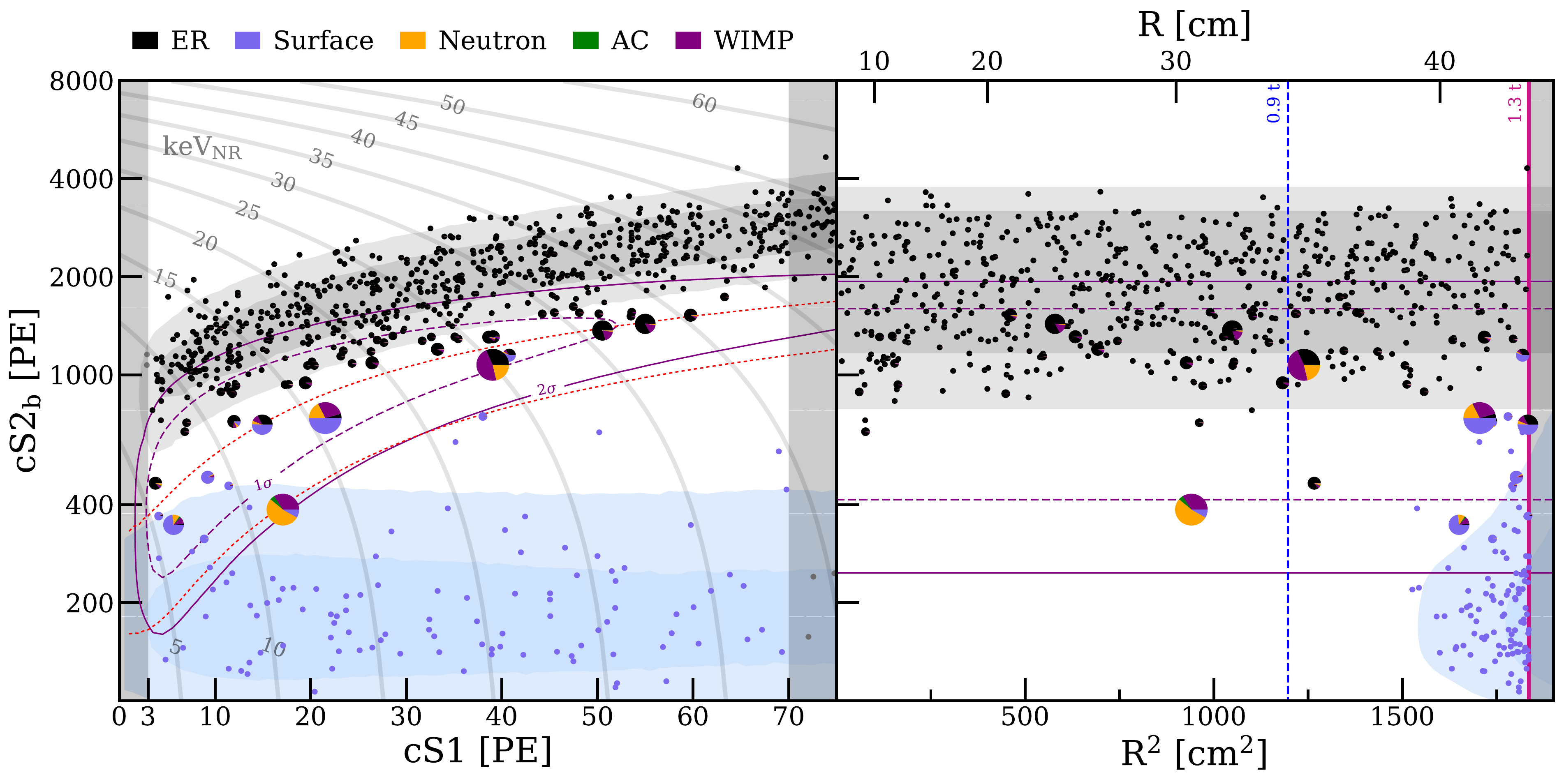}
\vspace{-0.1in}
\caption{\label{fig:datamc}DM search data in the \nnominalfiducialmassname \FM distributed in (\cSone,~\cStwob) (left) and (R$^2$,~\cStwob) (right) parameter spaces with the same marker descriptions as in \figref{fig:position_plots}. Shaded regions are similar to \figref{fig:position_plots}, showing the projections in each space of the surface (blue) and ER (gray) background components for \SRone. The $1\sigma$ (purple dashed) and $2\sigma$ (purple solid) percentiles of a 200~\gevcsq WIMP signal are overlaid for reference. Vertical shaded regions are outside the ROI. The NR signal reference region (left, between the two red dotted lines) and the maximum radii (right) of the \nonetonnemassname (blue dashed) and \nnominalfiducialmassname (magenta solid) masses are shown.  Gray lines show iso-energy contours in NR energy. 
}
\end{figure*}
 

The DM search data was blinded (\SRzero re-blinded after the publication of~\cite{xe_sr0}) 
in the signal region above the S2 threshold of 200~PE
and below the ER $-2\sigma$ quantile in (\cSone,~\cStwob) space,
prior to the tuning and development of event selection criteria and 
signal and background models.
Data quality criteria are imposed to include only well-reconstructed events and
to suppress known backgrounds. All events must contain a valid S1 and S2 pair.
S1s are required to
contain coincident signals from at least 3~PMTs within 100~ns.
The energy region of interest (ROI) is defined by \cSone between 3 and 70~PE,
corresponding to an average \searchregionkev (ER energy) or \searchregionkevnr (NR energy).
Furthermore, in order to suppress low-energy
accidental coincidence (\AC) events, S1 candidates must not have shape
properties compatible with S2 signals produced by single electrons.
The resulting S1 detection efficiency, estimated by simulation, is shown in \figref{fig:acc} and is smaller than that in~\cite{xe_sr0}
due to a wider S1 shape in the simulation
tuned to \krm and \rnzero data as well as properly accounting for mis-classification as S2.
This efficiency is consistent with that obtained by a data-driven method
where small S1s are simulated via bootstrapping PMT hits from 20-100 PE S1s. 

The signal ratio between the top and bottom PMT arrays is dependent on the depth at which the light is produced. For an S1 at a given interaction position, a 
p-value is computed based
on the observed and expected top/bottom ratio and p-values $<0.001$ are rejected. S2s are produced at the liquid-gas
interface and thus must have a compatible fraction of light seen in the top
array of $\sim$63\%. To reject events
coming from occasional light emission from malfunctioning PMTs, a threshold is placed on the maximum fractional 
contribution of a single PMT to an S1 signal.

The likelihoods of both the S1 and S2 observed hit-patterns compared to those
expected from simulation, given the reconstructed position, are used to reject events
that may be a result of multiple-scatters or \AC. 
The low-\cStwob, \cSone~=~68~PE, event found in~\cite{xe_sr0} did not pass event selection criteria in this analysis due to improvements to the MC simulation used for the S2 hit-pattern likelihood. To suppress events with poorly reconstructed hit-patterns that occur in regions with a 
high density of inactive PMT channels, the difference between the neural network and a likelihood-fit 
algorithm is required to be less than~2~to~5~cm, 
tighter towards larger S2 where fluctuations become negligible.
As in \cite{xe_sr0}, the width of the S2 signal in time must be compatible with the depth of the interaction, and the multiplicity of S1 and S2 signals must be consistent with a single-scatter event. 
The efficiency of all selection conditions is shown in \figref{fig:acc}, estimated using a combination of simulations and calibration control samples.

\begin{figure}[tbp]
\centering
\includegraphics[width=\columnwidth]{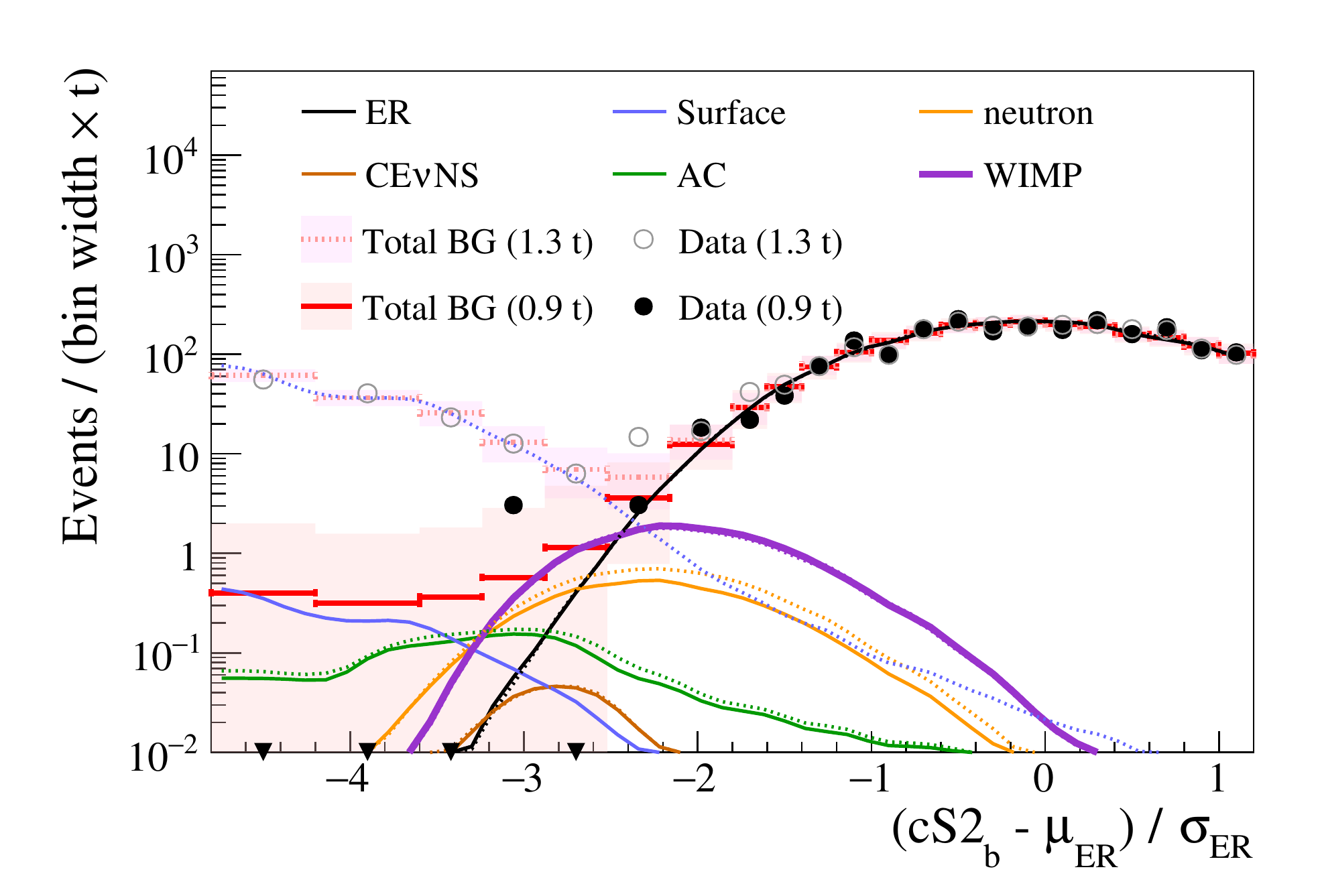}
\caption{\label{fig:datamc_proj}Background and 200~\gevcsq WIMP signal best-fit predictions, assuming $\sigma_{SI}$~=~\nbestfitsigmasitwoh, compared to DM search data in the \nonetonnemassname (solid lines and markers) and \nnominalfiducialmassname (dotted lines and hollow markers) masses. The horizontal axis is the projection along the ER mean ($\mu_{\mathrm{ER}}$), shown in \figref{fig:datamc}, normalized to the ER 1$\sigma$ quantile ($\sigma_{\mathrm{ER}}$). Shaded bands indicate the 68\% Poisson probability region for the total BG expectations. 
}
\end{figure}


This analysis expands on that in~\cite{xe_sr0} by modeling the radial distribution in the statistical inference procedure
and categorizing events at inner radii based on \z, such that the analysis space is \cSone, \cStwob, \radius, and \z. 
Each background component described below, and the WIMP NR signal, are modeled as a probability density function of all analysis dimensions.
For WIMP NR energy spectra, the Helm form factor for the nuclear cross section~\cite{lewin} and a standard isothermal DM halo as in~\cite{xe_sr0} are assumed, with $v_0 = 220\,\mathrm{km}/\mathrm{s}$, $\rho_\mathrm{DM} = 0.3~\mathrm{GeV}/(\mathrm{c}^{2}\times\mathrm{cm}^{3})$, $v_\mathrm{esc} = 544\,\mathrm{km}/\mathrm{s}$, and Earth velocity of $v_\mathrm{E} = 232\,\mathrm{km}/\mathrm{s}$.
These spectra are converted into the analysis space via the detector model described below. 
\Figsref{fig:position_plots}{fig:datamc} show the background and signal model shapes 
in various 2D projections of the analysis space compared to data. 
The 1D projection in \figref{fig:datamc_proj} and integrals in 
\tabref{table:bgmodel} show the absolute rate comparisons. 
An NR signal reference region is defined between 
the 200~\gevcsq WIMP median and $-2\sigma$ quantile in (\cSone,~\cStwob) space.

The ${}^{\mathrm{nat}}$Kr concentration in LXe is reduced via cryogenic distillation~\cite{xe_kr} to a sub-dominant level of
\nkrlevel, as determined from regular mass-spectrometry measurements~\cite{xe_sebastian}, and contributes an ER background rate of \krrate. 
The background contribution from the natural radioactivity of detector materials is suppressed within the fiducial volume to a similar level. Thus, the dominant ER background is 
from $\beta$-decays of $^{214}\mathrm{Pb}$ originating from $^{222}\mathrm{Rn}$ emanation.
The maximum and minimum decay rate of $^{214}\mathrm{Pb}$ is $(12.6\pm0.8)$ and $(5.1\pm0.5)\,\mu\rm{Bq/kg}$, estimated from $^{218}\mathrm{Po}$ $\alpha$-decays and time-coincident $^{214}\mathrm{Bi}$-$^{214}\mathrm{Po}$ decays, respectively, similarly to the method used in \cite{xe_rn}. The corresponding event rates in the ROI are 
$(71~\pm~8)$ and $(29~\pm~4)$~\dru.
The total ER background rate is stable throughout both science runs and measured as \nlowestbg after correcting for efficiency, which is the lowest background achieved in a dark matter detector to date.

The NR background includes contributions from radiogenic neutrons
originating from detector materials, coherent elastic neutrino nucleus scattering
(CE$\nu$NS) mainly from $^8$B solar neutrinos, and cosmogenic neutrons from secondary
particles produced by muon showers outside the TPC (negligible due to the muon veto~\cite{xe_muonveto}). The 
CE$\nu$NS rate is constrained by $^8$B solar neutrino
flux~\cite{solarflux} and cross-section~\cite{coherent}
measurements. The rate of radiogenic neutrons is
modeled with \textsc{Geant4} MC~\cite{geant4, xenon1t_mc_paper} using the measured radioactivity of materials~\cite{xe_screen}, assuming a normalization uncertainty of 50\% based
on the uncertainty in the \textsc{Sources\,4A}~\cite{sources4a} code and the difference between the \textsc{Geant4} and \textsc{MCNP} particle propagation simulation codes~\cite{mcnpx}. 
Fast neutrons have a mean free path of $\sim$15~cm in LXe and produce $\sim$5 times more multiple-scatter than single-scatter events in the detector, allowing for background suppression.
A dedicated search for multiple-scatter events finds 9 neutron candidates, consistent with the expectation of ($6.4\pm3.2$) derived
from the \textsc{Geant4} and detector response simulation described below, which is used to further constrain the expected single-scatter neutron event rate in DM search data.

The detector response to ERs and NRs is modeled similarly to the method described in Refs.~\cite{xe_sr0, xe100_tritium}. 
All \rnzero, \ambe, and neutron generator calibration data from both science runs are simultaneously
fitted to account for correlations of model parameters across different sources and runs.
To fit the \rnzero data, the parameterization of the ER recombination model is improved from~\cite{xe_sr0}
by modifying the Thomas-Imel model~\cite{ti_model}. These modifications include a power law field-dependence similar 
to~\cite{NEST_v1} to account for the different drift fields in each science run, an exponential energy dependence to extend 
the applicability to high-energy (up to $\sim$20~$\kevee$), and an empirical energy-dependent Fermi-Dirac suppression of the recombination 
at low-energy ($\lesssim2~\kevee$). 
The resulting light and charge yields after fitting are consistent 
with measurements~\cite{xe100_tritium, lux_tritium, pixey_ar37, lux_xe127}.
The fit posterior is used to predict the ER and NR 
distributions in the analysis space of the DM search data, 
achieving an ER rejection of \nerrejectionpct in the signal reference region,
as shown in \tabref{table:bgmodel}.
ER uncertainties in (\cSone,~\cStwob) are propagated for statistical inference via variation of 
the recombination and its fluctuation, 
as these show the most dominant effect on sensitivity (here defined as the median of an ensemble of confidence intervals derived under the background-only hypothesis~\cite{feldmancousins, PDG}). 
For WIMP signals, the uncertainties from all modeled processes are propagated into an uncertainty of 15\% (3\%) on the total efficiency
for 6~(200)~\gevcsq WIMPs. 

\begin{table}[ht]
\caption{Best-fit, including a 200~\gevcsq WIMP signal plus background,  expected event counts with \nlivetimetotal livetime in the \nnominalfiducialmassname \FM, \nonetonnemassname reference mass, and \ncoremassname core mass, for the full (\cSone,~\cStwob) ROI and, for illustration, in the NR signal reference region. The table lists each background (BG) component separately and in total, as well as the expectation for the WIMP signal assuming the best-fit $\sigma_{SI}$~=~\nbestfitsigmasitwoh. The observed events from data are also shown for comparison. Although the number of events in the reference region in the \nnominalfiducialmassname fiducial mass indicate an excess compared to the background expectation, the likelihood analysis, which considers both the full parameter space and the event distribution finds no significant WIMP-like contribution.
}
\centering
\newcolumntype{R}{>{\raggedleft\arraybackslash}X}
\bgroup
\def\arraystretch{1.2}
\begin{tabularx}{\columnwidth}{l R R R R}
\hline \hline
Mass & \nnominalfiducialmassname & \nnominalfiducialmassname & \nonetonnemassname & \ncoremassname \\
(\cSone, \cStwob) & Full & Reference & Reference & Reference \\
\hline 
ER &
    627$\pm$18 & 
    1.62$\pm$0.30 & 
    1.12$\pm$0.21 & 
    0.60$\pm$0.13 
    \\

neutron &
    1.43$\pm$0.66 & 
    0.77$\pm$0.35 & 
    0.41$\pm$0.19 & 
    0.14$\pm$0.07  
    \\

CE$\nu$NS &
    0.05$\pm$0.01 & 
    0.03$\pm$0.01 & 
    0.02 & 
    0.01 
    \\

AC &
    0.47$\substack{+0.27 \\ -0.00}$ & 
    0.10$\substack{+0.06 \\ -0.00}$ & 
   0.06$\substack{+0.03 \\ -0.00}$ & 
    0.04$\substack{+0.02 \\ -0.00}$ 
    \\

Surface &
    106$\pm$8 & 
    4.84$\pm$0.40 & 
    0.02& 
    0.01 
    \\

\hline
Total BG &
    735$\pm$20 & 
    7.36$\pm$0.61 & 
    1.62$\pm$0.28 & 
    0.80$\pm$0.14  
    \\
WIMP$_{\textrm{best-fit}}$ &
    3.56 & 
    1.70 & 
    1.16 & 
    0.83  
    \\
\hline
Data &
    739 & 
    14 &
    2 &
    2
    \\
\hline \hline
\vspace{0.001ex}
\end{tabularx}
\egroup
\raggedright
\label{table:bgmodel}
\end{table}

Energy deposits in charge- or light-insensitive regions produce lone S1s or S2s, respectively,
that may accidentally coincide and mimic a real interaction. 
The lone-S1 spectrum is derived from S1s occurring before the main S1 in high energy events and has a rate of [0.7,~1.1]~Hz. The uncertainty range is determined from differing 
rates of single electron S2s and dark counts in the time window before the event. 
The lone-S2 sample is composed of all triggered low-energy events containing S2s without a validly paired S1 and has a rate of $(2.6\pm0.1)$~mHz (without requiring the S2 threshold). The \AC background rate and distribution are estimated by randomly pairing lone-S1s and -S2s and simulating the necessary quantities for applying the event selection defined above. 

\rntwo progeny plate-out on the inner surface of the PTFE panels may decay 
and contaminate the search region if the reconstructed position falls within the \FM, herein referred to as ``surface'' background. 
Decays from $^{210}\mathrm{Pb}$ and its daughters that occur directly on the surface of the PTFE exhibit charge-loss 
due to S2 electrons being trapped on the surface and produce an S2/S1 ratio compatible with NR,
as shown in \figref{fig:datamc}. 
Several control samples are selected to derive a data-driven surface background model: DM search data reconstructed outside the TPC radius (due to position resolution) are used to predict the distribution in (\cSone, \cStwob, \z) via a kernel density estimator; the reconstructed \radius distribution of surface events depends only on the size of S2 and is modeled by fitting to a control sample composed of $^{210}\mathrm{Po}$ events as well as surface events with abnormally small S2/S1 from \rnzero calibration and DM search data (\cSone$>$~200~PE).
The (\radiussquared,~\cStwob) projection in \figref{fig:datamc} shows a correlation that provides 
additional discrimination power in the likelihood analysis.
Uncertainties in the radial shape are estimated by varying fitting methods. 
The normalization of the surface background is constrained 
by the bulk of surface events in DM search data shown in \figref{fig:datamc}.

The fiducial mass, 
shown as a magenta line in \figref{fig:position_plots}, 
is 8.0~cm below the liquid level to avoid mis-reconstructed
interactions in the GXe and 2.9~cm above the cathode to avoid interactions
in this region with a larger and less-uniform electric field. The corners of the \FM
are restricted further by requiring that the predicted total background rate in the ROI is flat to $<10\%$ in \z across slices of \radius,
such that the contribution from radioimpurities in detector materials to the ER background is sub-dominant 
relative to the uniform internal $^{214}$Pb contribution. 
The maximum radius (42.8~cm) was chosen to expect $\lesssim100$ surface-like events from the background model, to avoid over-constraining the corresponding tail prediction with these bulk events (\figref{fig:datamc}, right). 
This \FM contains \nnominalfiducialmass of LXe, determined from 
the total target mass of \nxenontargetmass and the fraction of \krm events contained inside.
An inner region containing \nonetonnemassname mass with \radius$<34.6$~cm is shown as a blue line in \figref{fig:position_plots} and is used to illustrate a reference region with negligible surface background rate.
Neutron interactions in the \FM occur mainly at extreme \z near 
the gate electrode or cathode as shown in \figref{fig:position_plots}, 
while WIMP NRs are expected to be uniformly distributed. This prompted designation of a \ncoremassname core mass, marked in green in \figref{fig:position_plots}, which contains a significantly lower neutron rate. \tabref{table:bgmodel} shows the number of events predicted in these regions by the post-fit models as well as the number of observed events after unblinding.

The data in the whole \nnominalfiducialmassname is interpreted using an unbinned extended likelihood with profiling over nuisance parameters \cite{James:1980ci, Bartlett:1953}. 
Modeling the surface background and adding the \radius dimension allows for the expansion of the 1~\ton~\FM in~\cite{xe_sr0} to \nnominalfiducialmassname, resulting in a 10\% sensitivity gain.
In addition to the three unbinned analysis dimensions (\cSone,~\cStwob,~\radius), events are categorized as being inside or outside the core mass (dependent on (\radius, \z).
All model uncertainties described above are included in the likelihood as nuisance parameters. 
A mis-modeling ``safeguard''~\cite{safeguard} (WIMP-like component) is added to the ER model and constrained by the \rnzero calibration data. This term prevents over- or under-estimation of the ER model in the signal region due to modeling choices. The ``anomalous leakage'' background component used in~\cite{xe_sr0} is not supported by the high statistics \rnzero calibration data in \SRone and is no longer included. \SRzero and \SRone are simultaneously fitted by assuming only the following 
parameters are correlated: electron-ion recombination in ER, neutron rate, WIMP mass and $\sigma_{SI}$. The best-fit in \figref{fig:datamc_proj} and \tabref{table:bgmodel} refers to the set of parameters that maximizes the likelihood.

Confidence intervals ($90\%\,\mathrm{C.L.}$) for ($\sigma_{SI}$,~mass) space were calculated by ``profile construction''~\cite{PDG, feldmancousins} using MC simulations and the coverage was tested for different values of nuisance parameters. This unifies one- and two-sided confidence interval constructions and avoids undercoverage that can result from applying asymptotic assumptions (Wilks' theorem). This asymptotic assumption was applied in the analysis of~\cite{xe_sr0} and caused a $\sim$38\% (44\%) decrease in the upper limit (median sensitivity) at a WIMP mass of 50~\gevcsq. 
A pre-unblinding decision was made to only report two-sided intervals if the detection significance exceeds $3\sigma$, which leads the analysis to conservatively overcover signals below the 68\% sensitivity band at 50~\gevcsq.
Two independent inference codes were developed and used to cross-check the result.


Event reconstruction and selection criteria were fixed prior to unblinding. However, four aspects of the models and 
statistical inference were modified after unblinding \cSone$<80$~PE, which together caused a 2\% (4\%) increase in the final limit (median sensitivity). First, the ER recombination parameterization, previously described, contains improvements implemented to solve a mis-modeling of the ER background in the NR ROI. The pre-unblinding parameterization included a sharp drop at $\sim$1.5~$\kevee$, which was sufficient for modeling the \SRzero \rnzero calibration data in~\cite{xe_sr0} but caused an enhancement to the safeguard term in a post-unblinding fit of the larger statistics \SRone \rnzero and DM search data.
The event at low-\z and low-\cStwob, indicated as mostly neutron-like in 
\figsref{fig:position_plots}{fig:datamc}, motivated scrutiny of the neutron model.
The second modification improved this model to correctly describe events with enlarged S1s due to additional scatters in the charge-insensitive region below the cathode. These events comprise 13\% of the total neutron rate in \tabref{table:bgmodel}. 
Third, we implemented the core mass segmentation to better reflect our knowledge of the neutron background's \z distribution, motivated again by the neutron-like event.
This shifts the probability of a neutron (50~\gevcsq WIMP) interpretation for this event in the best-fit model from 35\% (49\%) to 75\% (7\%) and improves the limit (median sensitivity) by 13\% (4\%). Fourth, the estimated signal efficiency decreased relative to the pre-unblinding model due to further matching of the simulated S1 waveform shape to \rnzero data, smaller uncertainties from improved understanding and treatment of detector systematics, and correction of an error in the S1 detection efficiency nuisance parameter. This latter set of improvements was not influenced by unblinded DM search data.

\begin{figure}
\includegraphics[width=\columnwidth]{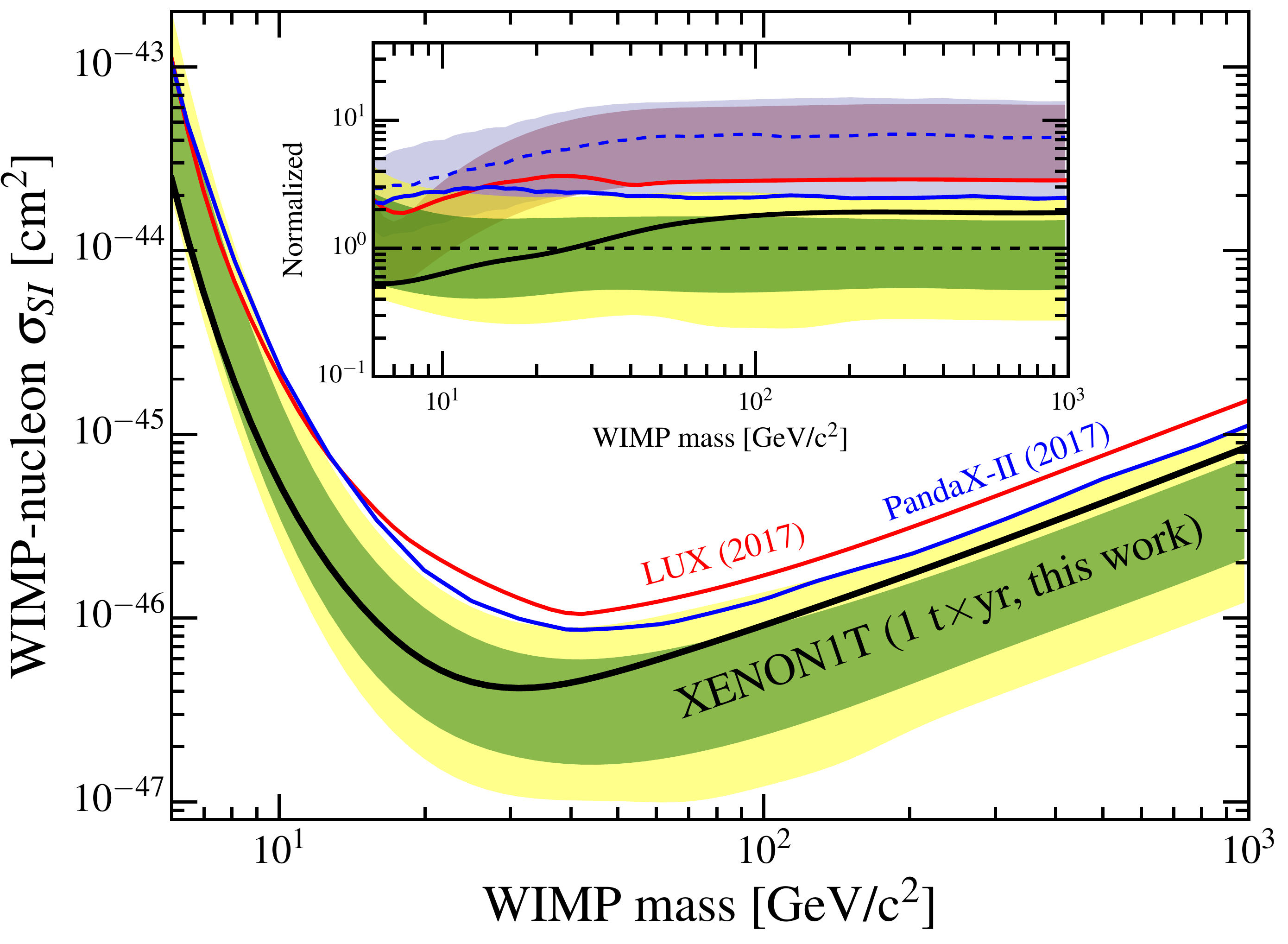}
\centering
\caption{\label{fig:contour} 90\% confidence level upper limit on $\sigma_{SI}$ from this work (thick black line) with the $1\sigma$ (green) and $2\sigma$ (yellow) sensitivity bands. Previous results from LUX~\cite{lux_dm} and \mbox{PandaX-II}~\cite{panda_dm} are shown for comparison. The inset shows these limits and corresponding $\pm1\sigma$ bands normalized to the median of this work's sensitivity band. The normalized median of the PandaX-II sensitivity band is shown as a dotted line. }
\end{figure} 

In addition to blinding, the data were also ``salted'' by injecting an undisclosed number and class of events in order to protect against fine-tuning of models or selection conditions in the post-unblinding phase. After the post-unblinding modifications described above, the number of injected salt and their properties were revealed to be two randomly selected \ambe events, which had not motivated any post-unblinding scrutiny. The number of events in the NR reference region
in \tabref{table:bgmodel} is consistent with background expectations. The profile likelihood analysis indicates no significant excesses in the \nnominalfiducialmassname \FM at any WIMP mass. A p-value calculation based on the likelihood ratio of the best-fit including signal to that of background-only gives $p=0.28$, 0.41, and~0.22 at 6,~50, and~200~\gevcsq WIMP masses, respectively. \Figref{fig:contour} shows the resulting 90\% confidence level upper limit on $\sigma_{SI}$, which falls within the predicted sensitivity range across all masses.
The $2\sigma$ sensitivity band spans an order of magnitude, indicating the large random variation in upper limits due to statistical fluctuations of the background (common to all rare-event searches). The sensitivity itself is unaffected by such fluctuations, and is thus the appropriate measure of the capabilities of an experiment~\cite{exo_nature}. The inset in \figref{fig:contour} shows that the median sensitivity of this search is $\sim$7.0 times better than previous experiments~\cite{lux_dm, panda_dm} at WIMP masses $>50$~\gevcsq.

\tabref{table:bgmodel} shows an excess in the data compared to the total background expectation in the reference region of the \nnominalfiducialmassname \FM. The background-only local p-value (based on Poisson statistics including a Gaussian uncertainty) is 0.03, which is not significant enough, including also an unknown trial factor, to trigger changes in the background model, fiducial boundary, or consideration of alternate signal models. This choice is conservative as it results in a weaker limit.


In summary, we performed a DM search using an exposure of \nlivetimetotal\,$\times$\,\nnominalfiducialmassname$=$~\nexposure, with an ER background rate of \nlowestbg, the lowest ever achieved in a DM search experiment. We found no significant excess above background and set an upper limit on 
the WIMP-nucleon spin-independent  
elastic scattering cross-section $\sigma_{SI}$ at \nmoststringentlimit for a mass of \nmoststingentmass, 
the most stringent limit to date for WIMP masses above 6~\gevcsq. An imminent detector upgrade, 
XENONnT, will increase the target mass to \nxenonnttargetmass. 
The sensitivity will improve upon this result by more than an order of magnitude.

We gratefully acknowledge support from the National Science Foundation, Swiss National Science Foundation, German Ministry for Education and Research, Max Planck Gesellschaft, Deutsche Forschungsgemeinschaft, Netherlands Organisation for Scientific Research (NWO), Netherlands eScience Center (NLeSC) with the support of the SURF Cooperative, Weizmann Institute of Science, Israeli Centers Of Research Excellence (I-CORE), Pazy-Vatat, Initial Training Network Invisibles (Marie Curie Actions, PITNGA-2011-289442), Fundacao para a Ciencia e a Tecnologia, Region des Pays de la Loire, Knut and Alice Wallenberg Foundation, Kavli Foundation, and Istituto Nazionale di Fisica Nucleare. Data processing is performed using infrastructures from the Open Science Grid and European Grid Initiative. We are grateful to Laboratori Nazionali del Gran Sasso for hosting and supporting the XENON project.

\end{document}